# Research topic trend prediction of scientific papers based on spatial enhancement and dynamic graph convolution network

Changwei Zheng, Zhe Xue*, Meiyu Liang and Feifei Kou

( *School of Computer Science (National Pilot School of Software Engineering), Beijing University of Posts and Telecommunications, Beijing Key Laboratory of Intelligent Telecommunication Software and Multimedia, Beijing 100876, China* )

**Abstract**
In recent years, with the increase of social investment in scientific research, the number of research results in various fields has increased significantly. Accurately and effectively predicting the trends of future research topics can help researchers discover future research hotspots. However, due to the increasingly close correlation between various research themes, there is a certain dependency relationship between a large number of research themes. Viewing a single research theme in isolation and using traditional sequence problem processing methods cannot effectively explore the spatial dependencies between these research themes. To simultaneously capture the spatial dependencies and temporal changes between research topics, we propose a deep neural network-based research topic hotness prediction algorithm, a spatiotemporal convolutional network model. Our model combines a graph convolutional neural network (GCN) and Temporal Convolutional Network (TCN), specifically, GCNs are used to learn the spatial dependencies of research topics a and use space dependence to strengthen spatial characteristics. TCN is used to learn the dynamics of research topics' trends. Optimization is based on the calculation of weighted losses based on time distance. Compared with the current mainstream sequence prediction models and similar spatiotemporal models on the paper datasets, experiments show that, in research topic prediction tasks, our model can effectively capture spatiotemporal relationships and the predictions outperform state-of-art baselines.

CLC number TP391

In recent years, with the rapid development of academic research, all kinds of academic literature have shown explosive growth [1]. In the face of massive literature, if we can effectively mine the development trend of research topics, it can provide valuable reference for scientific and technological workers in the research direction.

However, forecasting the trends in research topics presents special problems and challenges. various academic achievements are growing rapidly in recent years. For example, deep learning [2][3] has become a hot field of artificial intelligence research, which has grown significantly in recent years. However, there are also a large number of traditional research topics that have been replaced by new technology due to their inapplicability to changes and poor results. For example, the research interest of some traditional methods in the field of artificial intelligence, such as genetic algorithms, has declined year by year. And each research topic does not exist in isolation. Not only do similar research topics have mutual influences, but with the gradual deepening of interdisciplinary research, the influence of many research topics in different disciplines is also more closely [4]. For example, with the rise of graph neural networks, The transportation network begins to use the graph neural network [5][6] to capture the spatial features, and the research on intelligent transportation has gradually become a hot spot. Image recognition and other technologies are widely used in transportation, medical and other disciplines. Therefore, simply using the traditional sequence model [7][8] prediction model cannot effectively model the dependencies of the research topics, and it is necessary to combine the spatial dependencies [9] of the research topics.

As a more complex relational data structure that can describe entities, graphs can effectively represent the

———
Received date : 2022-03-16 ; Revised date :
**Fund projects** : This work was supported by National Key R&D Program of China (2018YFB1402600), and by the National Natural Science Foundation of China (61802028, 61772083, 61877006, 62002027) .
**Corresponding author** : Zhe Xue ( xuezhe@bupt.edu.cn)

spatial dependencies of research topics, and the representation learning of graph states [10][11] has gradually attracted attention. In recent years, with the rise of deep learning, new breakthroughs have been made in graph learning (such as GCN [12], GAT [13], Walk Pooling [14]), and people have a great interest in graph representation learning. The above figure shows that the learning work is only concentrated on the static graph, that is, the node characteristics and graph structure of the graph will not change. However, in reality, there are a large number of dynamically changing graphs. For example, the transmission path of the virus will change over time, and the number of infections at various locations will also change. Reflected in the graph, the structural features and node features of the graph will change dynamically according to time. Effectively predicting the epidemic situation in various places plays a crucial role in virus protection [15]. In social networks [16][17], people's social relationships are also dynamically changing, and user behaviors are constantly changing, so the vector representation of users should be updated accordingly. Likewise, the citation network of scientific articles is constantly enriched due to the frequent publication of new works citing existing technologies. Therefore, the popularity and influence of some technologies change with time [18][19].

In order to effectively mine the dependencies of research topics and predict the trend of each research topic, we constructed a dynamic graph of the dependencies of research topics at different times according to the citation relationship of each paper in the research topic. Model dependencies on research topics. We propose a spatially enhanced and dynamic graph convolutional network model SDTGCN (Spatial Dependency enhanced Temporal Graph Convolution Network) to capture the spatiotemporal features of research topics. The main contributions of this paper include three aspects :

(1) We adopt GCN to capture the dependencies of such research topics, and combine the specificity of citation counts as weights to enhance the spatial representation with the spatial dependency features of research topics .

(2) In order to mine the temporal relationship, we use a temporal convolutional network to fit the changes in the number of citations and the total amount of achievements respectively. And according to the characteristics of TCN, the prediction loss of each time point is weighted and calculated to enhance the robustness of the graph convolutional network.

(3) We compare and analyze the SDTGCN model proposed in this paper with the current state-of-the-art models related to spatiotemporal feature extraction through a large number of experiments. The experimental results show that our proposed method achieves the best performance in the trend prediction task of the research topic

## 1 Related work

With the rapid development of deep learning, cyclic and recursive network structures have begun to replace traditional linear models such as (ARIMA) and are widely used in sequence problems. However, models such as RNN are difficult to extract long-distance features effectively. Qin et al. [20] proposed a two-stage attention-based recurrent neural network, and on the basis of the attention mechanism, the LSTM structure was used to extract the long and short distance features of the time series. LSTNet [21] employs Convolutional Neural Networks (CNN) and Recurrent Neural Networks (RNN) to extract short-term local dependency patterns among variables and discover long-term patterns of time series. The Transformer [22] network structure proposed by Google adopts a pure attention mechanism and has achieved significant improvement in most sequence problems, but there is a problem of not being able to highlight local information. Since in the Transformer, the token only reflects the distance relationship through the position information, and theoretically all its nodes are equal, so the local information cannot be reflected. R-Transformer [23] first uses RNN to model local information, and then input it into Transformer, which avoids the problem of equalization of local information and global information. However, these network structures are difficult to parallelize, and the amount of computation is large . With the widespread application of convolutional neural networks in the field of two-dimensional images [24], people began to apply convolutional networks to one-dimensional sequence problems. Bai et al. For the problem of distance features, the use of dilated convolution and fully convolutional networks significantly improves the receptive field of the network, and proves that compared with LSTM\GRU and other models, the efficiency and accuracy are improved. He Xiaowei et al. [26] combined the advantages of high prediction accuracy of LSTM and short prediction time of GRU to efficiently predict cloud computing resource load.

To address the representation of spatial features in prediction tasks, T-GCN [27] combines graph convolutional networks (GCNs) and gated recurrent units (GRUs). Among them, GCN is used to learn complex topological structures and capture spatial correlations; Gated Recurrent Unit (GRU) is used to learn dynamic changes of traffic data and capture temporal correlations. Then, the T-GCN model is used for traffic prediction based on the urban road network. A3TGCN [28] utilizes spatiotemporal attention mechanism to learn dynamic spatiotemporal correlations of traffic data and combines spatiotemporal convolution for traffic prediction. LRGCN [29] treats temporal dependencies between temporally adjacent graph snapshots as a special relation to memory, and uses relational GCNs to jointly handle temporal and inter- temporal relations. AGCRN [30] employs a Node Adaptive Parameter Learning (NAPL) module to capture node-specific patterns, and a Data Adaptive Graph Generation (DAGG) module to automatically infer the interdependencies among different flows. The GCN in GC-LSTM [31] is capable of node structure learning for each time-slipped network snapshot, while LSTM is responsible for the temporal feature learning of network snapshots. Yu Ruiyun et al. [32] modeled the correlation between future demand and space-time through DCN and LSTM, and modeled the demand to predict the regional ride demand. Li Huibo [33] used an autoencoder as a framework, in which the encoder first uses DNN [34] to aggregate neighborhood information to obtain low-dimensional feature vectors, then uses GRU network to extract node temporal information, and finally uses the decoder to reconstruct the adjacency matrix and convert it Comparing with the real graph to construct the loss. Sun Yijun et al. [35] designed an appropriate weighting function for the edge features composed of k-nearest neighbor graphs to weaken the interference of far points, relatively strengthen the features of near points, and use a symmetric function that combines maximum pooling and average pooling to compensate. Global information loss.

## 2 Research topic dynamic graph representation method and task description of scientific papers

Based on the collected dblp paper data set, we divide the data set according to time $\Delta = \{D^{(1)}, D^{(2)}, ..., D^{(T)}\}$, which $D^{(T)}$ represents the collection of all papers published in the T year, and further we extract all keywords as research topics, and record the research topics and papers at the same time. Relationships and the number of papers on the research topic each year. We denote the set of research topics in year T as $N^{(T)}$, $Y^{(T)}$ is the number of papers on each research topic in year T. For research topic n, we set a window size, denoted as w, and take the total number of papers of research topic n in the past years from T-w+1 to T as the feature of n. The data of each time point d are processed to obtain the following feature sequence $N = \{N^{(1)}, N^{(2)}, ..., N^{(T)}\}$. In order to establish the association between research topics, at time T, for the research topics $u$, $v$, $q_{uv}$ denote the total number of papers in the research topic v applied to the papers in the research topic u. From this we obtain the directed weighted graph matrix $A^{(T)}$ at time T, where $q_{uv} = A_{u,v}^{(T)}$. Fr all time points, we can also obtain the following sequence $A = \{A^{(1)}, A^{(2)}, ..., A^{(T)}\}$. We take the feature sequence of research topics obtained by the above method as the node sequence of the graph, and the constructed weighted graph matrix as the adjacency matrix of the graph to construct the graph $\mathcal{G} = \{G^{(1)}, G^{(2)}, ..., G^{(T)}\}$. The goal of our task is to utilize sequences of a certain length $\mathcal{G}_{sub} = \{G^{(i)}, G^{(i+1)}, ..., G^{(j)}\}$ which $\mathcal{G}_{sub} \in \mathcal{G}$ to predict the num or each research topic $Y^{(j+1)}$ in the next year.

## 3. Research topic trend prediction method based on spatial enhancement and dynamic graph convolutional network

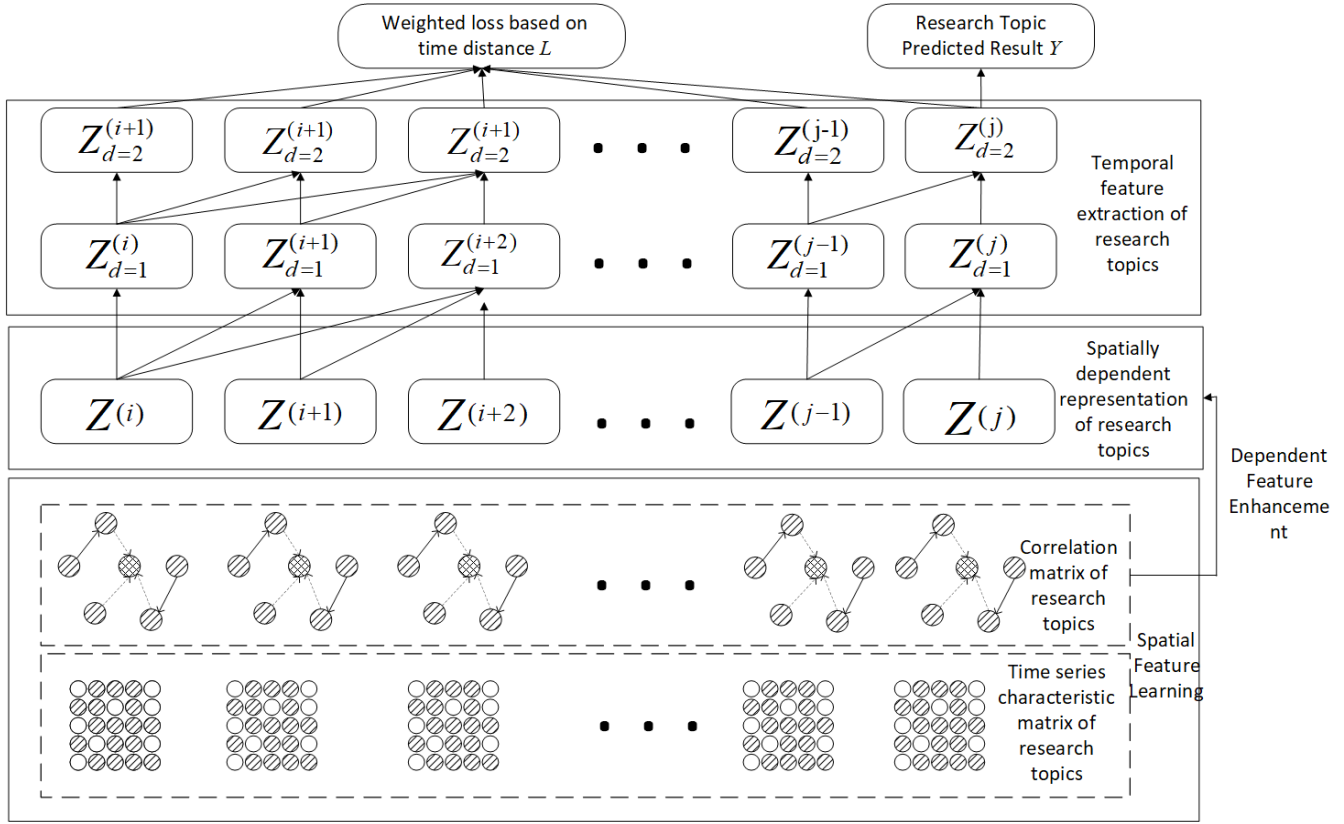

Fig. 1 Research topic trend prediction method based on spatial augmentation and dynamic graph convolutional network

In this section, we propose a convolutional network structure based on spatial enhancement and dynamic graph to simultaneously extract the spatial and temporal features of the dynamic graph of the research topic . The structure of the model is shown in Figure 1. In the extraction of temporal features, we combine the temporal features (total amount) of research topics and edge weight features (the number of citations between two research topics, and our graph structure changes with time).

3.1 Representation learning of research topic space dependence of scientific papers based on citation network spatial feature enhancement

For feature extraction of graphs, we use GCNs to capture spatial correlations. The convolution operation of a graph can be well approximated by a first-order Chebyshev polynomial expansion and generalized to a high-dimensional GCN as:

$$H = \left(I_N + D^{-\frac{1}{2}} A D^{-\frac{1}{2}}\right) X \Theta + b \quad (1)$$

where $A$ is the adjacency matrix of the graph at time point t, D is the degree matrix, $X$ is the feature matrix at time point t, and $\Theta \in R^{C\times F}$ and $b \in R^F$ denote learnable weight proofs and biases. In a dynamic graph, the structure of the graph changes at different points in time. If we learn a unique feature for each graph individually, the parameter size of the model will be too large and will lead to overfitting problems. We use the same graph convolutional network across the entire graphs, and the graph structure on all snapshots shares the same parameter matrix $\Theta$. According to the kipf setting, we use a two-layer GCN to aggregate node features of second-order neighbors, and use residual connections to connect the original features and the output of each layer.

However, GCN is mainly used in unweighted network structures, and for some conventional weighted graphs, the weights only measure the association between nodes for GCN, and the final node feature is equivalent to the weighting of other node features. beg for peace. For the research topic network we constructed, the weight between each research topic is not only a measure of the degree of dependence between two research topics, but also directly affects the number of results of the research topic. For example, for the research topic u, $\sum_{i=1}^{N} A_{li}^{t}$ Represents the time point t, the total number of papers cited by the research topic u, which is directly related to the number of papers on the research topic u, and finally the number of citations for each research topic

$$REF^t = \{\sum_{i=1}^{N} A_{1i}^t, \sum_{i=1}^{N} A_{2i}^t, \ldots \sum_{i=1}^{N} A_{Ni}^t\} \quad (2)$$

After normalization, the output spliced to the graph convolution is used for the extracted feature representation, and the output features of the final spatial representation learning are:

$$Z^t = concat(H, REF^t) \quad (3)$$

Where 'concat' represents the concatenation operation of the matrix.

3.2 Temporal convolution - based research topic trend prediction of scientific and technological papers

Among existing methods, GNN combined with recurrent neural networks such as LSTM and GRU has become a very popular network architecture for dynamic graph modeling. However, an increasing number of experimental and theoretical studies have shown that these RNNs cannot effectively capture long-range features. With the development of causal convolution and dilated convolution, the application of convolutional networks in time series tasks has gradually increased. To replace the traditional GRU model, inspired by TCN networks, we utilize convolutional networks to capture the temporal dependencies of dynamic graphs. For the research topic dependency graph at each time point, we use GCN to capture the spatial features to obtain the spatial feature representation of the research topic. These spatial features will be fed into the TCN. TCN uses causal convolution, i.e. a convolution where the output at time t is only convolved with earlier elements at time t and the previous layer. In order to improve the convolutional network, we use dilated convolution to improve the receptive field of the convolutional network. By stacking the convolutional network and setting an exponentially increasing expansion coefficient, the receptive field of the network is improved, and the computational complexity is reduced at the same time. The convolution process is as follows

$$Z(t) = \sum_{i=0}^{k-1} f(i) \cdot Z^{t-d \cdot i} \quad (4)$$

where k represents the size of the $f$ convolution kernel, the parameter of the convolution kernel, and d represents the expansion coefficient. We stack the depth of the network and adjust the expansion coefficient at each layer, which ensures that some filter will hit every input in the valid history, while also allowing the use of deep networks to capture long-range sequential features.

3.3 Weighted Loss Optimization for Temporal Distance-Based Research Theme Trend Prediction of Scientific Papers

Our model uses the same graph convolution parameters for fitting in extracting spatial dependencies. However, the spatial features captured by the graph convolution network indicate that after the graph convolution process, the features at different time points will have different weights, and often the most recent time point will have a higher weight, so that the graph convolutional network will be more inclined to learn the spatial features of the most recent time point, which is not conducive to learning the spatial structure of the entire time period. Therefore, we calculate the mean square error of each time point in the training set with its own next time point on the basis that the TCN network follows causal convolution:

$$loss_i = \|Y_i - \hat{Y}_i\| \quad (5)$$

where $Y_i$ represents the actual result and $\hat{Y}$ represents the predicted result. And give different weights according to the distance. When the time length of the training set is $t$, the weight of the i-th moment is i/t, and the final loss function is as follows

$$L = \sum_{i=1}^{t} (i/t) \|Y_i - \hat{Y}_i\| \quad (6)$$

The overall implementation steps of the algorithm are shown in the following table:

Table 1 Implementation steps of SDTGCN algorithm

| | |
|---|---|
| input | A collection of temporal characteristics of the research topic:<br>$$N = \{N^{(1)}, N^{(1)}, \ldots, N^{(T)}\}$$<br>A collection of time-series citation networks for research topics:<br>$$A = \{A^{(1)}, A^{(1)}, \ldots, A^{(T)}\}$$ |
| | 1. Extract a graph representation for each time point using a two-layer graph convolutional network<br>2. Use the original citation count network to calculate the citation count of the research topic, and calculate the citation count vector according to formula ( $REF^{(t)}$2)<br>3. according to Equation ( 3) to increase the spatial representation $Z^t$<br>4. According to formula ( 4) , the temporal features of the spatial representation sequence are captured by the temporal convolution module and the prediction results are obtained<br>5. According to formula ( 5)(6) , calculate the MSE error of each time point and its own true value at the next time point as the loss of model optimization |
| output | Prediction results of research topic trends at time $YT+1$<br>Weighted loss based on time distance $L$ |

## 4. Experiment results and analysis of the trend prediction of research topics of scientific and technological papers

4.1 Introduction to research topic datasets of scientific papers

  Dblp dataset We collected the paper data of dblp from 1970 to 2019 , and randomly selected 500 research topics from the top 20,000 research topics according to the number of papers on the research topics, and followed the above method. constructed dataset.

4.2 Experiment settings of research topic trend prediction of scientific and technological papers :

  In our experiments, we train a model with data from year i to i+T-1, and then use the model to predict the number of outcomes for each research topic in year T. We use 60% of the samples as the training set, 10% as the validation set, and 10% as the test set.

  To verify the effect of different receptive field sizes on each model, we set T to 8, 12, 16, and 20. We train different models for different T. Therefore, each model focuses on predicting the predicted value after a fixed time horizon, so the size of the training set increases with T. We train the model for up to 500 epochs. Starting from epoch 100, the model stops training if the loss does not improve on the validation set for 50 consecutive epochs. We use the Adam optimizer with a learning rate of 10-3. We set the number of hidden units in the graph convolutional layer to 32. Normalize and dropout the output of each GCN layer. The dropout of GCN and temporal convolutional network is set to 0.2. We store the model that achieved the highest validation accuracy and use it to make predictions on the test sample. For the temporal convolution module, we set the kernel size to 3.

  To evaluate the predictive performance of our method, we used 3 metrics to evaluate the difference between the true and predicted values $\hat{Y}_t$ , including mean absolute error (MAE) $Y_t$ , mean square error (MSE) , and variance fraction (Var) .

Specifically, MSE and MAE are used to measure prediction error: the smaller the predicted value, the better the prediction. Var calculates the correlation coefficient, which measures the ability of the prediction result to represent the actual data: the larger the prediction value, the better the prediction effect. For each experiment, we run 10 times and calculate the average as the result.

4.3 Introduction of experimental comparison method

  We compare the performance of the SDTGCN model with the following baseline methods: HA: Average the data from year i to i+T-1 as the prediction result for year T. LSTM: Use a two-layer LSTM network for prediction. A3TGCN: Graph Convolutional Network (GCN) and Gated Recurrent Unit (GRU) are combined to adjust the importance of different time points by adjusting the attention mechanism. GC-LSTM: A Graph Convolutional Network (GC) embedded Long Short-Term Memory (LTSM) network for end-to-end dynamic link prediction. LRGCN: It treats temporal dependencies between temporally adjacent graph snapshots as a special relationship with memory, and uses relational GCNs to jointly handle temporal and temporal relationships .

## 4.4 Analysis of the experimental results of the research theme trend prediction of scientific and technological papers

Table 2 Experimental results of the dblp paper dataset

| method name | index | 8 | 1 2 | 1 6 | 2 0 |
|---|---|---|---|---|---|
| HA | MAE | 1.1653 | 1.2815 | 1.3534 | 1.412 |
| | M SE | 2.0491 | 2.4262 | 2.6699 | 2.9352 |
| | VAR_ | -0.4695 | -0.4853 | -0.4408 | -0.3708 |
| L STM | MAE | 0.6891 | 0.6634 | 0.665 | 0.6293 |
| | M SE | 0.7177 | 0.6996 | 0 6 582 | 0.7747 |
| | VAR | 0. 176 | 0.1639 | 0.1 771 | 0.1795 |
| A 3TGCN | MAE | 0.6421 | 0.6074 | 0.6125 | 0.6 0 6 |
| | M SE | 0.7035 | 0.6595 | 0.6219 | 0.756 |
| | VAR_ | 0.1974 | 0.1892 | 0.2069 | 0.2203 |
| G CLST M | MAE | 0.5934 | 0.565 | 0.5651 | 0.6035 |
| | M SE | 0.6254 | 0.5871 | 0.5594 | 0.6568 |
| | VAR_ | 0.261 | 0.2407 | 0.2784 | 0.3121 |
| L RGCN | MAE | 0.6061 | 0.5728 | 0.5606 | 0.5917 |
| | M SE | 0.6526 | 0.5935 | 0.5493 | 0.6419 |
| | VAR_ | 0.2553 | 0.2497 | 0.2762 | 0.3293 |
| SDTGC N | MAE | 0.5435 | 0.4663 | 0.4627 | 0.4854 |
| | M SE | 0.5178 | 0.3713 | 0.3548 | 0.3963 |
| | VAR_ | 0.4523 | 0.5443 | 0.5409 | 0.5864 |

　　Table 2 shows the prediction results of our method on the paper research topic dataset, using time periods of length 8, 12, 16, and 20 to predict the next time points, respectively. It can be seen that this method achieves the best prediction performance under all evaluation metrics in all prediction ranges, proving the effectiveness of SDTGCN in spatiotemporal prediction. From the experimental results on the dataset of the research subject of the paper, we find that neural network-based methods, including SDTGCN model and LSTM model, which emphasize the importance of modeling temporal features, generally have better prediction accuracy than HA. For example, in the 8-layer prediction task, the MAE errors of SDTGCN and A3TGCN models are reduced by about 0.5232 and 0.6218, respectively, and the MSE is reduced by about 1.3456 and 1.5313 compared with HA. This is mainly due to methods such as HA, which find it difficult to handle complex, non-stationary time series data. And as the time length of training data increases, compared with the deep network model, the MAE and MSE error values of the shallow method both increase rapidly. In addition, the prediction accuracy of the methods based on spatiotemporal relationships (GCLSTM, LRGCN, A3TGCN) is higher than that of traditional sequence models. For example, when the time period length is 8, compared with the LSTM model, the MAE errors of the SDTGCN model and the A3TGCN model are reduced by about 0.1456 and 0.047, respectively, and the VAR of the SDTGCN and A3TGCN models are improved by 0.2763 and 0.021, respectively, compared with the TCN model, indicating that for For dynamic graph problems, spatial structure also plays an important role in prediction. Among methods that consider both spatial and temporal correlations, our method achieves significant improvements on all metrics. For example, when the time period length is 8, the MAE error of SDTGCN model is about 0.0986 and 0.0626 lower than that of A3TGCN and LRGCN models, respectively, and the accuracy is about 0.197 and 0.2549 higher than that of A3TGCN.

## 4.5 Analysis of the effect of sequence length on the prediction of SDTGCN and GCLSTM

　　In order to explore the fitting ability of our method to real data under different receptive fields, we conducted experiments with receptive fields of length 8, 12, 16, and 20 respectively , and the experimental results are shown in Table 2. Compared with the model GCLSTM , which utilizes lstm to capture temporal relationships , our model reduces the MAE error by about 0.0499, 0.0987, 0.1024, 0.1181 at receptive field lengths of 8, 12, 16, and 20, respectively. The accuracy is about 0.1913, 0.3036, 0.2625, 0.2743 higher than GCLSTM. For different receptive fields, our model has almost the same performance on every metric. For models such as gclstm and lrgcn , as the receptive field increases, the MSE and MAE indicators increase significantly, and the VAR indicator continues to decline. This shows that our model can still effectively capture long-range associations in long-term time series.

## 4.6 Comparison of prediction results of each module of SDTGCN

Table 3 Prediction experiment of each module of SDTGCN

| method name | | 8 | 12 | 16 | 20 |
|---|---|---|---|---|---|
| TCN | MAE | 0.6931 | 0.6669 | 0.648 | 0.6044 |
| | MSE | 0.704 | 0.6883 | 0.6644 | 0.6441 |
| | VAR | 0.1775 | 0.1748 | 0.1813 | 0.2031 |
| GCN | MAE | 0.647 | 0.6159 | 0.5606 | 0.5917 |
| | MSE | 0.7365 | 0.6832 | 0.5493 | 0.6419 |
| | VAR | 0.1978 | 0.1393 | 0.2762 | 0.3293 |
| SDTGCN | MAE | 0.5435 | 0.4663 | 0.4627 | 0.4854 |
| | MSE | 0.5178 | 0.3713 | 0.3548 | 0.3963 |
| | VAR | 0.4523 | 0.5443 | 0.5409 | 0.5864 |

　　In order to verify the role of each module of the model, we compared this method with TCN and GCN respectively. The experimental settings for each comparison method are as follows: For the TCN network, we remove the graph convolutional network and only use the node features as the input of the network. For GCN, we replace the temporal convolution module with a linear layer and take the output of the linear layer as the prediction result.

　　The experimental results are shown in Table 3. Compared with GCN, when the receptive fields are 8, 12, 16 and 20, respectively, for the indicator MSE, our model loss is reduced by 0.1035, 0.1496, 0.0979, 0.1063, and for MAE, our The model decreased by 0.2187, 0.3119, 0.1945, 0.2456, and on the VAR metric, our model increased by 0.2545, 0.405, 0.2647, 0.2571. These results show the importance of modeling time dependencies for these datasets. Our model also effectively simulates time. According to the MSE metric, when the receptive fields are 8, 12, 16, and 20, our model improves by 0.1496, 0.2006, 0.1853, 0.119 on the MAE metric compared to TCN, respectively. For MSE, our model drops by 0.1862, 0.317, 0.3096, 0.2478, respectively. For the indicator VAR, our model improves by 0.2748, 0.3695, 0.3596, 0.3833. This shows that our method effectively utilizes spatial features to enhance predictions on sequence data. This is because our method effectively utilizes the prediction results of the adjacency matrix and enhances the spatial features, thereby improving the overall prediction ability of our method for the future.

4.7 Performance analysis of spatiotemporal feature representation model

　　Model running efficiency comparison: We conducted 50 rounds of training for each model with the receptive field sizes of 8, 12, 16, and 20, respectively, and calculated the model running time. The results are shown in Figure 2. It can be found that our model takes the shortest time to run, and as the perception increases, the time-consuming of other models increases faster, while our model grows more steadily. For example, when the receptive field size is 8, the method in this paper The time-consuming of A3TGCN and A3TGCN are 2.4s and 2.8s, respectively. When the receptive field is 20, the time-consuming is 4.9s and 6.7s, respectively. These results show that our model is less computationally expensive than other comparative methods. Thanks to the hierarchical structure of the expanded convolutional network and the role of the expansion factor, our model can increase the receptive field exponentially by stacking the number of convolutional layers, and because the parameters of the convolutional network are generally smaller than the recurrent recursive unit parameters, so that the computational complexity of the model is significantly reduced.

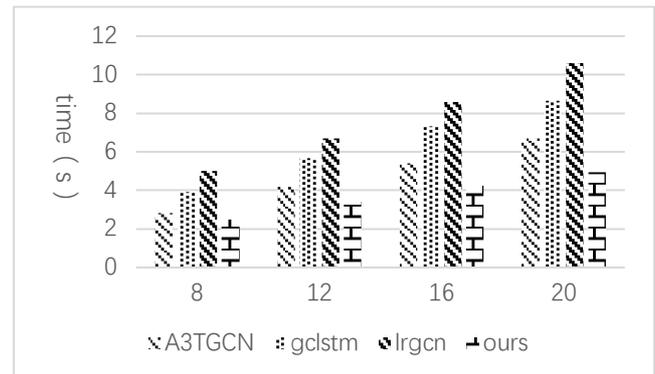

Fig. 2. Comparison results of model running time under different receptive fields

4.8 Analysis of the trend prediction results of research topics of scientific and technological papers

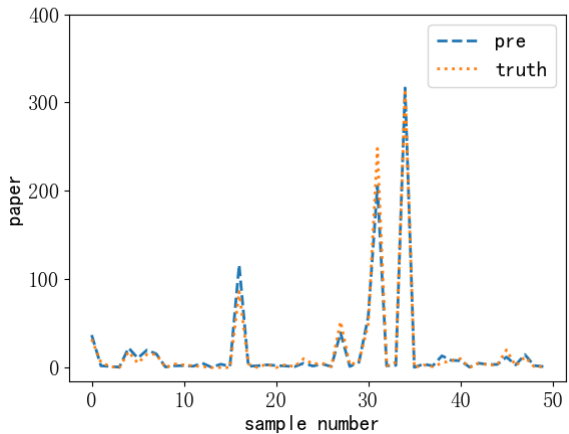

Fig. 3 Visualization of prediction results and true values

We randomly selected 50 research topics from 500 research topics, and selected a window size of 8 to compare the predicted results of SDTGCN with the real results. The experimental results are shown in Figure 3 . As can be seen from the figure, for most research topics, the prediction results of our model basically fit the true value. But for some peak parts, such as the prediction of the 31st and 34th samples, the prediction results of our method are 279.17 and 355.84 respectively , while the real values are 310 and 370 respectively. The prediction results of our method are the same as the real values . The values differ by about 30 and 15, respectively . The predicted results of our method are lower than the true results at the peak, which is due to the convolution of our network from the adjacent nodes in space and the adjacent time points. The convolution process is performed by the weighted sum of adjacent features. , to make the peaks smoother, especially for spatial features, the number of papers with some research topics is extremely low, and even zero in some years, so in the graph convolution operation, due to the aggregation of these nodes, resulting in The predicted value of its surrounding adjacent nodes is generally small. Combined with the influence of graph convolution and temporal convolution on feature extraction, the final prediction value for some spiked parts is small.

## 5 Conclusion

Aiming at the problem of forecasting the trend of research topics in scientific papers, this paper constructs the dependencies between research topics according to the citation relationship of the papers, thereby describing the spatial relationship between research topics. According to the citation relationship at different times and the changes in the number of papers of each research topic, according to the time dimension, a dynamic graph of the research topic relationship network is constructed to represent the characteristics of time series changes. A combined architecture of GCN network and TCN network is used to predict the trend of research topics, in which the GCN network is used to extract the spatial dependencies between research topics, and combine the influence of the number of citations of papers to enhance the spatial representation. The TCN network is used to capture temporal dependencies. By stacking dilated convolutional layers, the running time consumption of the model is effectively reduced, and the ability to capture long-distance dependencies is improved. In order to improve the learning ability of GCN for the research topic graph, we use a weighted loss to improve the fitting ability of the graph network to the graph structure at different time points, and optimize the model by minimizing this loss. Finally, the optimal prediction results are obtained in the research topic trend prediction task of scientific papers.

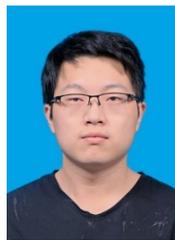

**Changwei Zheng,** born in 1997, master's degree. The main research directions are natural language processing, data mining, deep learning

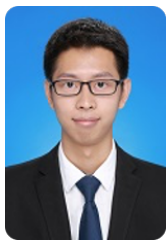

**Zhe Xue**, born in 1987, associate professor and master tutor. His main research interests are machine learning, artificial intelligence, data mining, and image processing. He has published more than 30 papers and 1 academic monograph. Presided over the National Natural Science Foundation of China Youth Fund project, participated in the national key research and development program projects, etc.

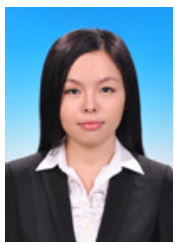

**Meiyu Liang,** born in 1985. Associate professor and master tutor of School of Computer Science(National Pilot School of Software Engineering), Beijing University of Posts and Telecommunications. The main research directions are artificial intelligence, data mining, multimedia information processing, computer vision, etc.

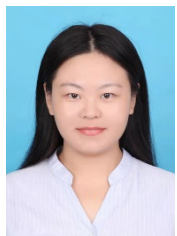

**Feifei Kou**, born in 1989. Lecturer of School of Computer Science (National Pilot School of Software Engineering), Beijing University of Posts and Telecommunications. Her research interests include semantic learning and multimedia information processing.